\documentclass[conference, 10pt]{IEEEtran}

\usepackage{latexsym,epsf,times,amsmath,color,amssymb,indentfirst,subfigure,fancyhdr,colortbl,bm,caption}
\usepackage{xspace}
\usepackage{graphicx}\epsfxsize=3.in
\usepackage{bbm}
\usepackage{cite}
\usepackage{algorithm}
\usepackage{algorithmic}
\usepackage{amssymb}
\usepackage{amsmath}
\usepackage{cite}
\usepackage{stfloats}

\ifCLASSINFOpdf

\else
 
\fi

\begin{document}
%
\title{Energy Efficient Precoder Design for MIMO-OFDM with Rate-dependent Circuit Power}

\author{\IEEEauthorblockN{Zijian Wang, Ivan Stupia, Luc Vandendorpe}
\IEEEauthorblockA{Institute of Information and Communication Technologies, Electronics and Applied Mathematics\\Universit\'{e} catholique de Louvain\\Place du Levant 2,
B-1348 Louvain-la-Neuve, Belgium\\\{first name.last name\}@uclouvain.be}
}

\maketitle

\begin{abstract}
This paper studies an energy efficient design of precoders for point-to-point multiple-input-multiple-output (MIMO) orthogonal frequency-division multiplexing (OFDM) systems. Differently from traditional approaches, the optimal power allocation strategy is studied by modelling the circuit power as a rate-dependent function. We show that if the circuit power is a constant plus an increasing and convex function of the transmission rate, the problem of minimizing the consumed energy per bit received can be reformulated as a convex fractional program and solved by means of a bisection algorithm. The impact of the some system parameters is investigated either analytically or by means of computational results.
\end{abstract}

\IEEEpeerreviewmaketitle

\section{Introduction}
Nowadays, energy efficiency for wireless communications is becoming a main economical and societal challenge~\cite{LXXYZCX11}. Hence, instead of maximizing the information rate under a certain power constraint, engineers wish to maximize the transmission rate per consumed power. This is also referred to as the energy efficiency (EE) maximization problem. Among the early contributions in this area, the authors in \cite{KD10} originally considered the transmission rate per consumed energy to adapt the transmission mode (i.e. number of streams, space time signaling, MIMO detection, etc.) of a MIMO-OFDM system. Paper~\cite{BL11} investigates the optimal precoding strategy for EE maximization in MIMO systems, showing that without considering the power used to feed circuitries, the optimum transmission power will be zero for a Gaussian MIMO channel. 
In~\cite{ICJF12}, the authors used the mathematical tool of convex fractional programming to address the power allocation problem for EE maximization in OFDM systems, showing the optimality of the waterfilling power allocation policy. More recently, the authors of \cite{HHJY14} extended this work to a multi-user and multi-cell scenario. 

An alternative formulation of the original EE maximization problem was proposed by the authors of~\cite{IF10} who stated it as the minimization of the consumed energy per bit of information received. This route has also been pursued in \cite{PD10} and \cite{ED12} to investigate the power allocation problem in OFDM systems, confirming the waterfilling nature of the optimal solution. 
All the papers above, however, consider the overall circuit power consumption (except the transmit power) to be either zero or a constant value. On the contrary, the authors of~\cite{WV14} accounted for the variability of the circuit power consumption for discrete-time base-band signal processing (due to different modulation orders and coding rates) by modelling this power as being dependent on the transmission rate. 

In the current work, we consider the EE design of precoders for MIMO-OFDM systems, formulated as the minimization of the consumed energy per bit of information received.  As a main result, capitalizing on the quasi-convexity of the EE function, we derive the optimum precoding and power allocation strategy. Eventually, the impact of system parameters is mathematically analyzed and the results are confirmed by means of computational results. 

In this paper, boldface lowercase letters and boldface uppercase
letters represent vectors and matrices, respectively. Notation $(\cdot)^H$ and $\texttt{tr}\left(\cdot\right)$ denote
 conjugate transpose operation and trace of a matrix respectively. $|\cdot|$ is the determinant of the matrix. $\left(\textbf{A}\right)_m$ denotes the $(m,m)$-th entry of the matrix $\textbf{A}$.  $\mathbf{I}_N$ represents the $N{\times}N$ identity
matrix. $\phi'(x)$ stands for the first-order derivative of the function $\phi(x)$. $X_{dB}$ is $X$ expressed in decibels. 
The notation $[x]_0^+$ denotes $\texttt{max}\left\{0, x\right\}$.
Finally, we denote the expectation operation by $\mathrm{E}\left\{\cdot\right\}$.

\section{system model}
Consider a MIMO-OFDM system with $M$ transmit antennas, $N$ receive antennas and $K$ subcarriers. Assuming perfect synchronization, the observation model at the $k$-th subcarrier is given by 
\begin{equation}
\textbf{y}_k=\sqrt{\beta}\textbf{H}_k\textbf{x}_k+\textbf{n}_k.
\end{equation}
$\textbf{H}_k\in \mathbb{C}^{N\times M}$ is the MIMO channel matrix for the $k$-th subcarrier, whose entries are assumed to be independent and identically distributed (i.i.d.) zero-mean  circularly-symmetric complex Gaussian random variables with unit variance. $\beta=\left(G_1M_l d^{n}\right)^{-1}$ is the path loss budget where $n$ is the path loss exponent, $G_1$ is the gain factor at $d=1$m and $M_l$ is the link margin accounting for the hardware process variations and other noise and interference \cite{CGB05}. $\textbf{x}_k$ is an $M\times 1$ vector which denotes the transmitted symbols with a transmission power $P_k={\rm E}\{\textbf{x}_k^H\textbf{x}_k\}$.
$\textbf{n}_k$ is a zero-mean complex additive white Gaussian noise
vector with covariance $\texttt{E}\left\{\textbf{n}_k\textbf{n}_k^H\right\}=\sigma^2\textbf{I}_N$, where $\sigma^2=BN_0N_f$, $B$ is the bandwidth per subcarrier, $N_0$ is the one-sided noise power spectral density,  and $N_f$ is the noise figure defined as in \cite{CGB05}.

According to \cite{WV14}, the total power consumption of the transceiver can be modeled as 
\begin{equation}
P_{\texttt{total}}=\frac{1}{\omega}\sum_{k=1}^K P_k+P_c+\kappa\phi\left(B\sum_{k=1}^K\theta_k\right)(Watt),
\end{equation}
where $\omega$ is the efficiency of the power amplifier, $P_c=\rho_{tc}M+\rho_{rc}N$ accounts for the power needed to feed the radio frequency (RF) chain~\cite{CGB05}, while $\kappa\phi\left(B\sum_{k=1}^K\theta_k\right)$ is the rate-dependent term due to the discrete-time baseband signal processing. $\kappa$ represents a constant coefficient, $\theta_k$ is the information rate at the $k$-th subcarrier, and 
 $\phi(\cdot)$ is assumed to be monotonically increasing function with $\phi(0)=0$.
Traditionally, the EE function is defined as the information rate divided by the total energy consumption:$\left(B\sum_{k=1}^K\theta_k\right)/{P_{\texttt{total}}}$. 
In this paper, we pursue a different route by investigating the consumed energy per bit of information received, i.e.
\begin{equation}
\varepsilon=\frac{\frac{1}{\omega}\sum_{k=1}^K P_k+P_c+\kappa\phi\left(B\sum_{k=1}^K\theta_k\right)}{B\sum_{k=1}^K\theta_k}(Joule/bit).
\label{eef}
\end{equation}
In the following, we will see that thanks to this formulation of the performance metric, we can restate the problem as a convex program and derive the optimal solution.

\section{Energy efficient precoder design}
In this section, we first restate the original fractional program as a convex optimization problem and then we derive the optimal precoding strategy. Finally we propose an algorithm achieving the minimum energy consumption per bit received.

\subsection{Precoder optimization}

Our aim is to minimize the function defined as in \eqref{eef}. 
The problem can therefore be formulated as
\begin{equation} \label{opt_prob}
\min_{\{\textbf{Q}_k\}}\frac{\frac{1}{\omega}\sum_{k=1}^K P_k(\textbf{Q}_k)+P_c+\kappa\phi\left(B\sum_{k=1}^K\theta_k(\textbf{Q}_k)\right)}{B\sum_{k=1}^K\theta_k(\textbf{Q}_k)}
\end{equation}
where $\textbf{Q}_k$ is the covariance transmission matrix for the $k$-th subcarrier with $P_k\left(\textbf{Q}_k\right) = \texttt{tr}\left(\textbf{Q}_k\right) = {\rm E} \{ \textbf{x}^H_k\textbf{x}_k \} $.
Assuming Gaussian codewords, we have
\begin{equation}\label{information}
\theta_k=\texttt{log}_2\left|\textbf{I}_N+\frac{\beta\textbf{H}_k\textbf{Q}_k\textbf{H}_k^H}{\sigma^2}\right|\left(bits/channel~use\right).
\end{equation}

We first show the following lemma.
\newtheorem{theorem}{Theorem}
\newtheorem{lemma}[theorem]{Lemma}
\begin{lemma}
The optimal structure of $\textbf{Q}_k$ is $\textbf{Q}_k=\textbf{V}_k\widetilde{\textbf{Q}}_k\textbf{V}_k^H$, $k=1,...,K$, where $\widetilde{\textbf{Q}}_k$ is a diagonal matrix and $\textbf{V}_k$ comes from the 
singular value decomposition (SVD) of $\textbf{H}_k$ that $\textbf{H}_k=\textbf{U}_k\mathbf{\Sigma}_k\textbf{V}_k^H$.
\end{lemma}
\begin{IEEEproof}
Assume that $\textbf{Q}_k^{\star}$ is the solution of \eqref{opt_prob} and  $\textbf{V}_k^H\textbf{Q}_k^{\star}\textbf{V}_k$ is not diagonal. According to SVD of $\textbf{H}_k$, we have 
\begin{equation}\label{svd}
\beta\textbf{H}_k^H\textbf{H}_k=\textbf{V}_k \mathbf{\Lambda}_k\textbf{V}_k^H,
\end{equation}
where $\textbf{U}_k$ is a unitary matrix and $\mathbf{\Lambda}_k=\beta \mathbf{\Sigma}_k^H\mathbf{\Sigma}_k$ is a diagonal matrix. 
Define $\widetilde{\textbf{Q}}_k^\star=\textbf{V}_k^H\textbf{Q}_k^\star\textbf{V}_k$,
where
$\texttt{tr}\left(\widetilde{\textbf{Q}}_k^\star\right)=\texttt{tr}\left(\textbf{Q}_k^\star\right)$,
 thanks to Hadamard's inequality, we have
\begin{align}
\texttt{log}_2\left|\textbf{I}_N+\frac{\beta\textbf{H}_k\textbf{Q}_k^{\star}\textbf{H}_k^H}{\sigma^2}\right|&=\texttt{log}_2\left|\textbf{I}_M+\frac{\widetilde{\textbf{Q}}_k^{\star}\mathbf{\Lambda}_k}{\sigma^2}\right|
\nonumber\\&<\texttt{log}_2\prod_{m=1}^M\left(1+\frac{(\widetilde{\textbf{Q}}_k^{\star}\mathbf{\Lambda}_k)_{m}}{\sigma^2}\right)\nonumber\\
&= \texttt{log}_2\prod_{m=1}^M\left(1+\frac{(\widetilde{\textbf{Q}}_k^{\star})_{m}(\mathbf{\Lambda}_k)_{m}}{\sigma^2}\right),
\end{align}
where the first equality is due to Sylvester's determinant theorem that $\texttt{log}_2\left|\textbf{I}_p+\textbf{A}\textbf{B}\right|=\texttt{log}_2\left|\textbf{I}_q+\textbf{B}\textbf{A}\right|$, where $\textbf{A}$ is a $p\times q$ matrix and $\textbf{B}$ is a $q\times p$ matrix.
Therefore there exists a diagonal matrix $\widetilde{\textbf{Q}}_k'$ such that 
\begin{equation}
P_k'=\texttt{tr}\left(\widetilde{\textbf{Q}}_k'\right)<\texttt{tr}\left(\widetilde{\textbf{Q}}_k^{{\star}}\right)=P_k^{{\star}},
\end{equation}
and
\begin{equation}
\texttt{log}_2\prod_{m=1}^M\left(1+\frac{(\widetilde{\textbf{Q}}_k')_{m}(\mathbf{\Lambda}_k)_{m}}{\sigma^2}\right)=\texttt{log}_2\left|\textbf{I}_N+\frac{\beta\textbf{H}_k\textbf{Q}_k^{\star}\textbf{H}_k^H}{\sigma^2}\right|.
\end{equation}
Then from (\ref{eef}), defining $\textbf{Q}_k'=\textbf{U}_k^H\widetilde{\textbf{Q}}_k'\textbf{U}_k$, we have 
\begin{equation}
\varepsilon\left\{\textbf{Q}_1^{\star},...,\textbf{Q}_k',...,\textbf{Q}_K^{\star}\right\}<\varepsilon\left\{\textbf{Q}_1^{\star},...,\textbf{Q}_k^{{\star}},...,\textbf{Q}_K^{\star}\right\},
\end{equation}
which is contradictory with the fact that $\textbf{Q}_k^{\star}$ minimizes the objective function.
Therefore, every matrix $\widetilde{\textbf{Q}}_k^\star$, $k=1,...,K$, should be a diagonal matrix.
\end{IEEEproof}
In the following, we only consider the subset of matrices $\textbf{Q}_k$ that have the structure as in Lemma 1. Hence, the rate for subcarrier $k$ can be rewritten as
\begin{align}
\theta_k & = \texttt{log}_2\left|\textbf{I}_N+\frac{\beta\textbf{H}_k\textbf{Q}_k\textbf{H}_k^H}{\sigma^2}\right| \nonumber \\
 & = \sum_{m=1}^M\texttt{log}_2\left(1+\frac{\left(\widetilde{\textbf{Q}}_k\right)_{m}\left(\mathbf{\Lambda}_k\right)_{m}}{\sigma^2}\right)\stackrel{\Delta}{=}\sum_{m=1}^M \theta_{k,m},
\end{align}
which defines $\theta_{k,m}$.
As
\begin{equation}
\left(\widetilde{\textbf{Q}}_k\right)_{m}=\frac{\sigma^2}{\left(\mathbf{\Lambda}_k\right)_{m}}\left(2^{\theta_{k,m}}-1\right),
\label{ctoq}
\end{equation}
and
\begin{equation}
P_k=\sum_{m=1}^M \left(\widetilde{\textbf{Q}}_k\right)_{m},
\end{equation}
it appears that there is a one-to-one mapping between the rate and the covariance matrix, and that $P_k$ can be written as a function of $\theta_{k,1},...,\theta_{k,M}$.
Therefore, instead of formulating the problem in the variables $\textbf{Q}_k$,  
we can restate it in the variables $\theta_{k,m}$, which is similar to the approach of \cite{IF10}:
\begin{align}
\min_{\{\theta_{k,m}\}, t\in \mathbb{R}_{+}} & t\left({\frac{1}{\omega}\sum\limits_{k=1}^K P_k+P_c+\kappa\phi\left(B\sum\limits_{k=1}^K\sum\limits_{m=1}^M\theta_{k,m}\right)}\right) \nonumber \\
s.t.  \quad &t(B\sum\limits_{k=1}^K \sum\limits_{m=1}^M\theta_{k,m})=1
\end{align}  
where, for a given $t$, this problem is convex. The corresponding Lagrangian is:
\begin{align}
&L\left(\theta_{1,1}, \cdots, \theta_{K,M}, \upsilon\right)\nonumber\\&=t\left(P_c+\frac{1}{\omega}\sum_{k=1}^K\sum_{m=1}^M \frac{\sigma^2}{\left(\mathbf{\Lambda}_k\right)_{m}}\left(2^{\theta_{k,m}}-1\right)+\kappa\phi\left(\frac{1}{t}\right)\right)\nonumber\\
&+\upsilon \left(1-tB\sum_{k=1}^K\sum_{m=1}^M \theta_{k,m}\right).
\end{align}
Therefore, the optimal solution must fulfill the following KKT conditions:
\begin{align}
\theta_{k,m}^{\star}&\geq 0,\quad k=1,...,K\\
1-tB\sum_{k=1}^K\sum_{m=1}^M \theta_{k,m}^{\star}&=0\\
\frac{t}{\omega}\cdot\frac{\sigma^2\ln 2}{\left(\mathbf{\Lambda}_k\right)_{m}}2^{\theta_{k,m}^{\star}}-\upsilon tB&=0,\quad k=1,...,K \label{kkt}
\end{align}
which, after some manipulation of (\ref{kkt}), leads to 
\begin{equation} \label{eq:theta}
\theta_{k,m}^{\star}=\left[\texttt{log}_2 \left(\upsilon B \omega\right)-\texttt{log}_2\left(\frac{\sigma^2 \ln 2}{\left(\mathbf{\Lambda}_k\right)_{m}}\right)\right]_0^+.
\end{equation}
Substituting \eqref{eq:theta} into \eqref{ctoq} get 
\begin{equation}
(\widetilde{\textbf{Q}}_k)_{m}^{\star}=\left[\frac{\omega \upsilon B}{\ln 2}-\frac{\sigma^2}{(\mathbf{\Lambda}_k)_{m}}\right]_0^+.\label{waterlevel}
\end{equation}
Defining the water level as being $\mu=\frac{\omega \upsilon B}{\ln 2}$, the next step is to find its optimal value. 
This can be obtained by rewriting the total power ($P=\sum_{k=1}^K P_k$) as a function of the total rate ($\Theta=\sum_{k=1}^K \theta_k$), leading to
\begin{equation}\label{p_r}
P(\Theta)=\sigma^2\sum_{l=1}^L\left(\sqrt[L]{\frac{2^\Theta}{\prod{\Lambda}_l}}-\frac{1}{{\Lambda}_l}\right),
\end{equation}
because
\begin{equation}\label{r_w}
\Theta=\sum_{l=1}^L\texttt{log}_2\left(1+\frac{p_l{\Lambda}_l}{\sigma^2}\right)=\sum_{l=1}^L\texttt{log}_2\left(\frac{\mu{\Lambda}_l}{\sigma^2}\right),
\end{equation}
where $L$ is the number of eigenchannels receiving a non-zero power among the $\min \{M, N\} \times K$ space/frequency subchannels. $p_l$ denotes the transmission power and ${\Lambda}_l$ denotes the channel gain for the $l$-th subchannel.
The derivative of the transmit power with respect to the total rate is given by
\begin{equation}
\frac{dP}{d\Theta}=\frac{ \sigma^2 \ln 2}{\sqrt[L]{\prod\limits_{l'=1}^L{\Lambda}_{l'}}}\cdot 2^{\frac{\Theta}{L}},
\label{mimod}
\end{equation}
from which it can clearly be seen that $P(\Theta)$ is a strictly increasing and strictly convex function.

Then we have the following lemma:
\begin{lemma}
If $\frac{P(\Theta)}{\omega}+\kappa\phi(B\Theta)$ is convex, then $\varepsilon(\Theta)$ is quasi-convex.
\end{lemma}
\begin{IEEEproof}
Let us denote $g(\Theta) = \frac{P(\Theta)}{\omega}+\kappa\phi(B\Theta)$. We have 
\begin{equation}
\varepsilon(\Theta)=\frac{P_c+g(\Theta)}{B\Theta}.
\end{equation}
Assume that $\Theta_1<\Theta_2$ and $0<\lambda<1$. If 
\begin{equation} \label{quasi_convexity1}
\varepsilon\left(\lambda\Theta_1+(1-\lambda)\Theta_2\right)>\varepsilon(\Theta_1),
\end{equation}
we have 
\begin{equation}
\frac{P_c+g\left(\lambda\Theta_1+(1-\lambda)\Theta_2\right)}{B\left(\lambda\Theta_1+(1-\lambda)\Theta_2\right)}>\frac{P_c+g(\Theta_1)}{B \; \Theta_1 }.
\end{equation}
Noting that, thanks to the convexity of $g(\Theta)$, 
\begin{equation}
g\left(\lambda\Theta_1+(1-\lambda)\Theta_2\right)<\lambda g\left(\Theta_1\right)+\left(1-\lambda\right) g\left(\Theta_2\right),
\end{equation}
we get
\begin{align}
&P_c B \left(\lambda\Theta_1+(1-\lambda)\Theta_2\right)+Bg(\Theta_1)\left(\lambda\Theta_1+(1-\lambda)\Theta_2\right)\nonumber\\
&<P_c B \Theta_1+B\Theta_1 g\left(\lambda\Theta_1+(1-\lambda)\Theta_2\right)\nonumber\\
&<P_c B \Theta_1+B\Theta_1 \lambda g\left(\Theta_1\right)+B\Theta_1\left(1-\lambda\right) g\left(\Theta_2\right)
\end{align}
which after some manipulations, leads to
\begin{equation}
P_c<\frac{\Theta_1g(\Theta_2)-\Theta_2g(\Theta_1)}{\Theta_2-\Theta_1}.
\label{meet1}
\end{equation}
If instead of \eqref{quasi_convexity1}, we assume that
\begin{equation}
\varepsilon\left(\lambda\Theta_1+(1-\lambda)\Theta_2\right)>\varepsilon(\Theta_2),
\end{equation}
similarly we get
\begin{equation}
P_c>\frac{\Theta_1g(\Theta_2)-\Theta_2g(\Theta_1)}{\Theta_2-\Theta_1}.
\label{meet2}
\end{equation}
Equations (\ref{meet1}) and (\ref{meet2}) cannot be met simultaneously, which means that 
\begin{equation}
\varepsilon\left(\lambda\Theta_1+(1-\lambda)\Theta_2\right)\leq\texttt{max}\left\{\varepsilon(\Theta_1),\varepsilon(\Theta_2)\right\},
\end{equation}
which implies that $\varepsilon(\Theta)$ is a quasi-convex function.
\end{IEEEproof}
Lemma~2 means that  $\frac{P(\Theta)}{\omega}+\kappa\phi(B\Theta)$ should be convex to guarantee that the objective function is quasi-convex.

The optimal transmission rate should fulfill the following condition:
\begin{equation}
\varepsilon'(\Theta)=\frac{1}{B\Theta}\left(P_{\texttt{total}}'(\Theta)-\frac{P_\texttt{total}(\Theta)}{\Theta}\right)=0.
\label{DEZ}
\end{equation}

Let us consider a generic convex model for the baseband power consumption, in the form of $\phi(B\Theta)=(B\Theta)^{\alpha}$ ($\alpha\ge1$). Then, we obtain
\begin{align}
P_{\texttt{total}}'(\Theta) & =\alpha \kappa B^{\alpha}\Theta^{\alpha-1}+\frac{P'(\Theta)}{\omega} \nonumber \\ 
 & = \alpha \kappa B^{\alpha}\Theta^{\alpha-1}+\frac{\mu(\Theta)\texttt{ln}2}{\omega},
\end{align}
where we made explicit the dependence of $\mu$ with respect to $\Theta$.
Hence, the value of the total information rate minimizing the consumed energy per bit must be such that
\begin{align}
\varepsilon'(\Theta) =\frac{\mu(\Theta)\texttt{ln}2}{\omega} & -\frac{1}{\Theta}\left(\frac{P(\Theta)}{\omega}+P_c\right) \nonumber \\ 
 & +\left(\alpha-1\right)\kappa B^\alpha \Theta^{\alpha-1} = 0. \label{key2}
\end{align}

\subsection{Algorithm description}

In the previous subsection, we showed that the original fractional program can be solved by simply finding an information rate value that satisfies \eqref{key2}. 
Unfortunately, such a value cannot be obtained in a closed form. Hence, we resort to an iterative approach based on a bisection algorithm. 
The algorithm is provided hereafter. The initial value $\Theta_u^{0}$ can be set to any arbitrary strictly positive value. The iteration stops when the search interval becomes smaller than a prescribed threshold $\delta$.

\begin{table}[ht]
\begin{algorithmic} [1]
\STATE Set $\Theta_l=0$; $\Theta_u=\Theta_u^{0}$
\STATE Calculate $\mu_u$ and $P_u$ at the point $\Theta_u$. Calculate $\varepsilon'(\Theta_u)$.
\WHILE{$\varepsilon'(\Theta_u)<0$}
\STATE $\Theta_u=\Theta_u\times 2$
\ENDWHILE
\WHILE{$\Theta_u-\Theta_l>\delta$}
\STATE $\Theta_c=0.5(\Theta_l+\Theta_u)$
\STATE Calculate $\Pi(\Theta_c)$
\IF {$\varepsilon'(\Theta_c)=0$}
\STATE go to line 15
\ELSIF {$\Pi(\Theta_c)>0$}
\STATE $\Theta_u=\Theta_c$
\ELSE
\STATE $\Theta_l=\Theta_c$
\ENDIF
\ENDWHILE
\end{algorithmic}
\caption{Algorithm description}
\end{table}

\section{Impact of system parameters}
In this section, we analyze the impact of some system parameters on the EE link performance metric. We first consider the impact of the distance $d$ on the minimum consumed energy per received bit when all the other parameters are kept to a given constant value. 

Assume two different values of the transmitter-receiver separation $d$, namely $d_1$ and $d_2$, where $d_2>d_1$. Denote by $\varepsilon(\Theta;d)$ the consumed energy per bit evaluated for a rate $\Theta$ at distance $d$. Denote by $P_c(d)$ the power consumption needed to feed the RF chains for distance $d$ and by
\begin{align}
\Theta^\star(d) = \arg \min_{\Theta} \{\varepsilon(\Theta;d)\}  
\end{align} 
the value of the rate $\Theta$ corresponding to the minimum consumed energy per bit at distance $d$. 
Finally, denote by $P(\Theta;d)$ the total transmit power evaluated at distance $d$ to achieve the rate $\Theta$. 
Then we have
\begin{align}
&\varepsilon(\Theta^{\star}({d_2});{d_2})\nonumber\\
&=\frac{\frac{P(\Theta^{\star}({d_2}),d_2)}{\omega}+P_c({d_2})+\kappa\phi\left(B\Theta^{\star}({d_2})\right)}{B\Theta^{\star}({d_2})}\nonumber
\\&=\frac{\left(\frac{d_2}{d_1}\right)^n\cdot\frac{P(\Theta^{\star}({d_2}),d_1)}{\omega}+P_c({d_1})+\kappa\phi\left(B\Theta^{\star}({d_2})\right)}{B\Theta^{\star}({d_2})}\nonumber
\\&>\frac{\frac{P(\Theta^{\star}({d_2}),d_1)}{\omega}+P_c({d_1})+\kappa\phi\left(B\Theta^{\star}({d_2})\right)}{B\Theta^{\star}({d_2})}\nonumber
\\&=\varepsilon(\Theta^{\star}({d_2});{d_1})\geq\varepsilon(\Theta^{\star}({d_1});{d_1}).
\label{ana2}
\end{align}
The second equality comes from (\ref{svd}), (\ref{p_r}) and \eqref{r_w} observing that, when the distance $d$ increases from $d_1$ to $d_2$, the channel gain of every subchannel ($\Lambda_l$ in \eqref{p_r}) is multiplied by $(d_2/d_1)^{-n}$. Hence, according to \eqref{r_w}, the rate can be kept constant if the water level is multiplied by $(d_2/d_1)^{n}$. Since the rate is a strictly monotonic function of the water level, this is the unique possible choice. 
Thus according to the waterfilling policy, the number of subchannels with non-zero power is not changed, and the total transmit power is multiplied by $(d_2/d_1)^{n}$. $P_c$ is independent of the value $d$. The last inequality is due to the fact that $\Theta^{\star}({d_1})$ minimizes the function when $d=d_1$. Therefore we have, $\varepsilon(\Theta^{\star}({d_2});{d_2})>\varepsilon(\Theta^{\star}({d_1});{d_1})$, meaning that the minimum number of Joules per received bit increases with the distance.


Let us now analyze the impact of distance $d$ on the rate value at the optimal point. Let us define $P_\texttt{total}(\Theta,d)$ as the total power written as a function of $\Theta$ and $d$. 
With the definitions provided above, we have the following theorem:
\begin{theorem} \label{th3}
If $\phi(\cdot)$ is linear, then $\Theta^{\star}(d_2)<\Theta^{\star}(d_1)$ for $d_2>d_1$.
\end{theorem}
\begin{IEEEproof}
Since $\Theta^{\star}(d_1)$ is the optimal point that minimizes the objective function when $d=d_1$, according to (\ref{DEZ}),
\begin{align}
\varepsilon'(\Theta^{\star}(d_1),d_1)&=\frac{P_{\texttt{total}}'(\Theta^{\star}(d_1),d_1)-\frac{P_\texttt{total}(\Theta^{\star}(d_1),d_1)}{\Theta^{\star}(d_1)}}{B\Theta^{\star}(d_1)}
\nonumber\\
&=0.
\end{align}
If $\phi(\cdot)$ is linear, after some manipulations, we get
\begin{equation}
\frac{P'(\Theta^{\star}(d_1),d_1)}{\omega}=\frac{\frac{P(\Theta^{\star}(d_1),d_1)}{\omega}+P_c(d_1)}{\Theta^{\star}(d_1)}.
\end{equation}
Then 
\begin{align}
\varepsilon'(\Theta^{\star}(d_1),d_2)
&=\frac{1}{B\Theta^{\star}(d_1)}\left(\frac{P'(\Theta^{\star}(d_1),{d_2})}{\omega}\right. \nonumber
\\
&\left.
-\frac{1}{\Theta^{\star}(d_1)}\left(\frac{P(\Theta^{\star}(d_1),{d_2})}{\omega}+P_c(d_1)\right)\right)
\nonumber
\\
&= \frac{1}{B\Theta^{\star}(d_1)}\left(\left(\frac{d_2}{d_1}\right)^n\cdot\frac{P'(\Theta^{\star}(d_1),{d_1})}{\omega}\right.
\nonumber\\
&\left.
-\frac{1}{\Theta^{\star}(d_1)}\left(\left(\frac{d_2}{d_1}\right)^n\cdot\frac{P(\Theta^{\star}(d_1),{d_1})}{\omega}+P_c\right)\right) \nonumber \\
&=\frac{1}{B\Theta^{\star}(d_1)}\left(\left(\frac{d_2}{d_1}\right)^n\cdot\frac{P_c(d_1)}{\Theta^{\star}(d_1)}-\frac{P_c(d_1)}{\Theta^{\star}(d_1)}\right)
\nonumber \\
&>0.
\end{align}
where the derivative of the power in the first line together with equation \eqref{mimod} yields to the third line.
Using the fact that $\varepsilon(\Theta,d)$ is quasi-convex, $\varepsilon'(\Theta,d)>0$ for $\Theta>\Theta^{\star}(d)$, $\varepsilon'(\Theta,d)<0$ for $\Theta<\Theta^{\star}(d)$, which is proven in detail in Theorem 1 in~\cite{WV14}. Therefore
we have $\Theta^{\star}(d_1)>\Theta^{\star}(d_2)$.
\end{IEEEproof}

A similar analysis can be carried out for parameters $\sigma^2$, $\rho_{tc}$ and $\rho_{rc}$. We however omit it for the sake of concision. It appears that increasing the noise variance $\sigma^2$ has an impact similar to increasing the distance, meaning that the optimal rate value decreases with increasing noise variance. As far as $\rho_{\rm tc}$ and $\rho_{\rm rc}$ are concerned, increasing their value also leads to a decrease of the corresponding optimal value of the rate.

\section{Numerical results}
In this section, we illustrate our analytical findings by means of numerical results. The system parameters are set as follows:
the bandwidth for each subcarrier is set to $B=10$KHz, $\rho_{tc}=82.5$mW, $\rho_{rc}=105.5$mW~\cite{CGB05}, and we fix $\kappa=5\times 10^{-8}$ as in \cite{WV14}. We also select the following values: $n=3.5$,  ${G_0}_{dB}=-\left({G_1}_{dB}+{M_l}_{dB}\right)=-70$dB where ${G_1}_{dB}=30$dB is the gain factor at $d=1$m and ${M_l}_{dB}=40$dB  \cite{CGB05}. The noise power spectral density is set to ${N_0}_{dB}=-170$dBm/Hz, the noise figure to ${N_f}_{dB}=10$dB/Hz and the amplifier efficiency is chosen to be $\omega=0.4$. Unless otherwise specified, we use a parameter $\alpha=1$ for the rate dependent power consumption term. Finally, we set $\delta=0.01$ as the tolerance of the bisection algorithm.
The results are averaged over 1000 different channel realizations.

\begin{figure}
\centering
\includegraphics[width=3.5in]{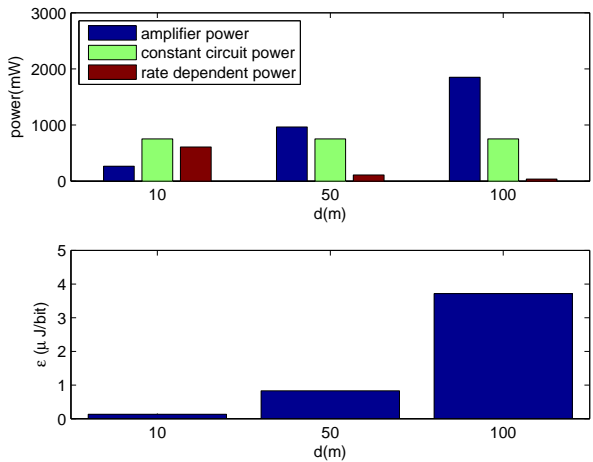}
\caption{Power consumption and EE comparison for different $d$ with $M=N=4$ and $K=64$.}
\label{f2}
\end{figure}                                                 
\begin{figure}
\centering
\includegraphics[width=3.5in]{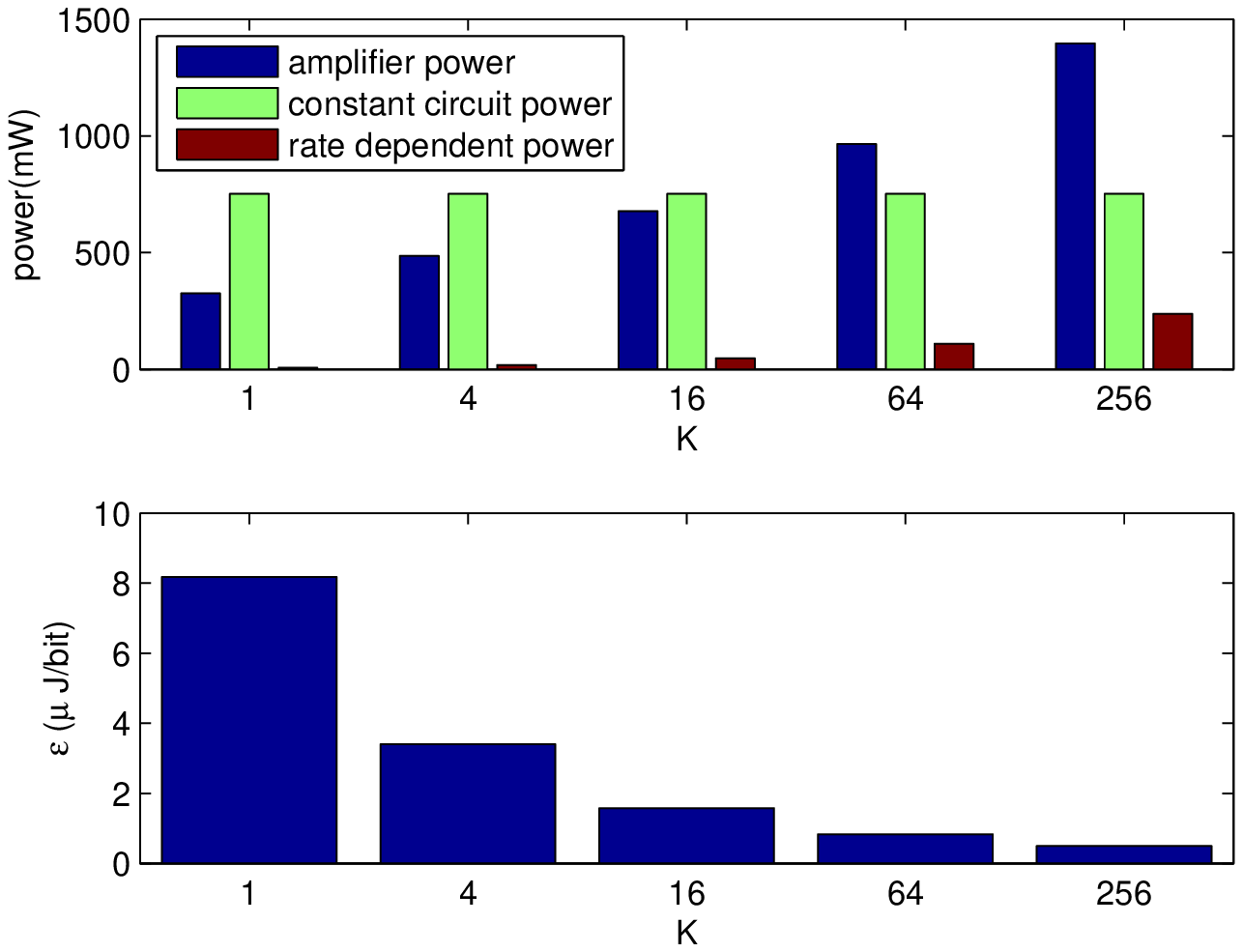}
\caption{Power consumption and EE comparison for different $K$ with $M=N=4$ and $d$=50m.}
\label{f1}
\end{figure}

\subsection{Effect of link distance $d$}
Fig.\ref{f2} reports both the different power consumption terms and the value of the objective function $\varepsilon$ at the optimal point, for different values of $d=10, 50, 100$m. 
The number of antennas and the number of carriers are set to $M=N=4$ and $K=64$, respectively. As said before, the value of the objective function $\varepsilon$ increases with the distance. So does the transmit power. As expected from Theorem \ref{th3}, the optimum rate value decreases with increasing distance which means that the rate dependent power consumption also decreases with increasing distance. In view of these different dependencies of power consumption terms with respect to distance it is however difficult to predict the evolution of the total power with the distance. 

\subsection{Effect of bandwidth enlargement}
Fig.\ref{f1} reports both the different power consumption terms and the value of the objective function $\varepsilon$ at the optimal point, for a number of subcarriers $K$ growing from $1$ to $256$, meaning different bandwidth sizes $KB$. We select $d=50$m and $M=N=4$. As it is expected, it can be observed that the consumed energy per bit decreases when the number of subcarriers grows. As a matter of fact, when the number of carriers is increased from $K_1$ to $K_2$ the solution space for $K_1$ is contained in the solution space of $K_2$ and the solution for $K=K_2$ cannot be worse than the solution for $K=K_1$. It is also observed that both the transmit power and the rate dependent power grow with $K$, meaning that the rate increases with an increasing value of $K$. 
\begin{figure}
\centering
\includegraphics[width=3.5in]{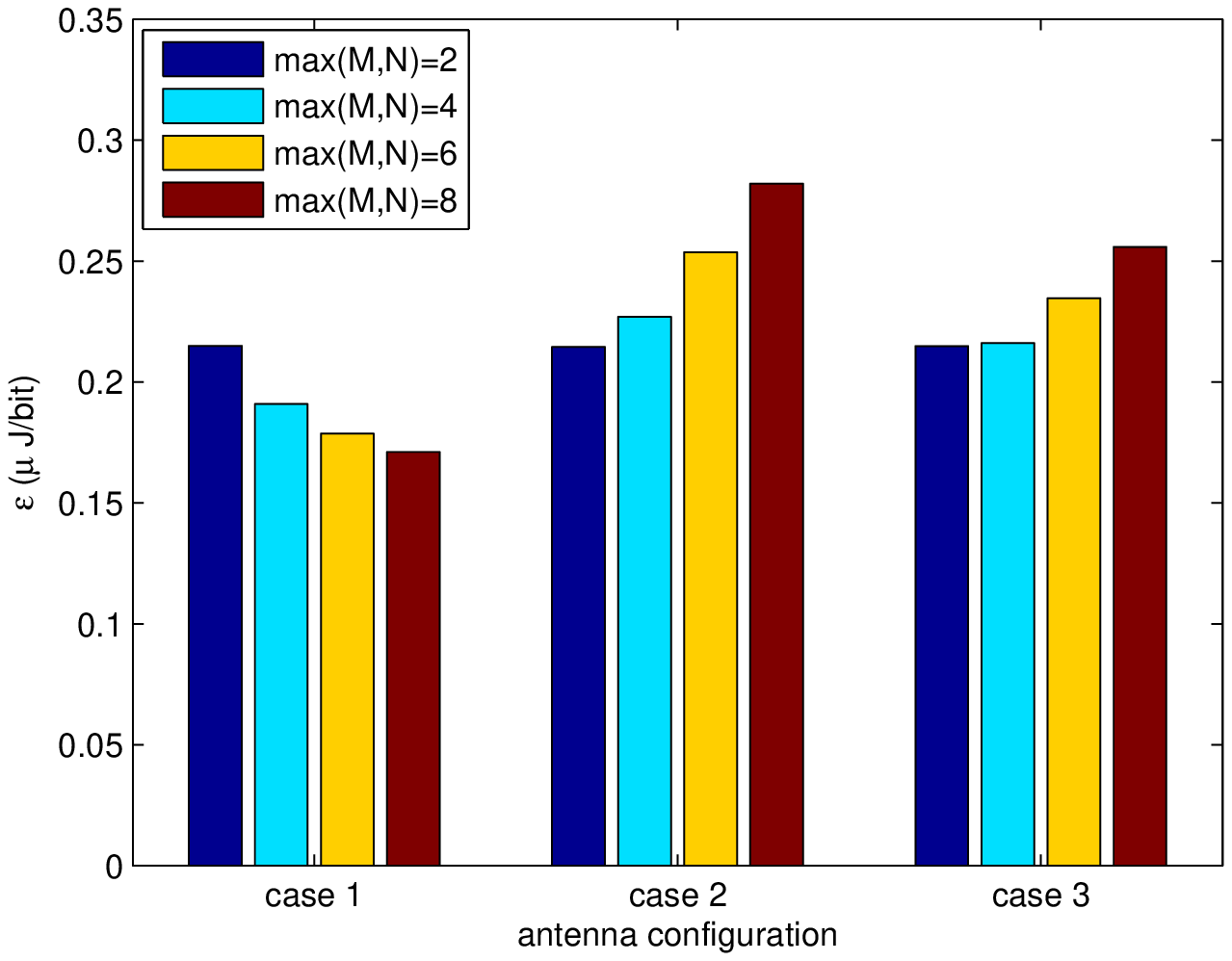}
\caption{Power consumption and EE comparison for different antenna configurations with $d$=10m and $K=32$.}
\label{f3}
\end{figure}
\begin{figure}
\centering
\includegraphics[width=3.5in]{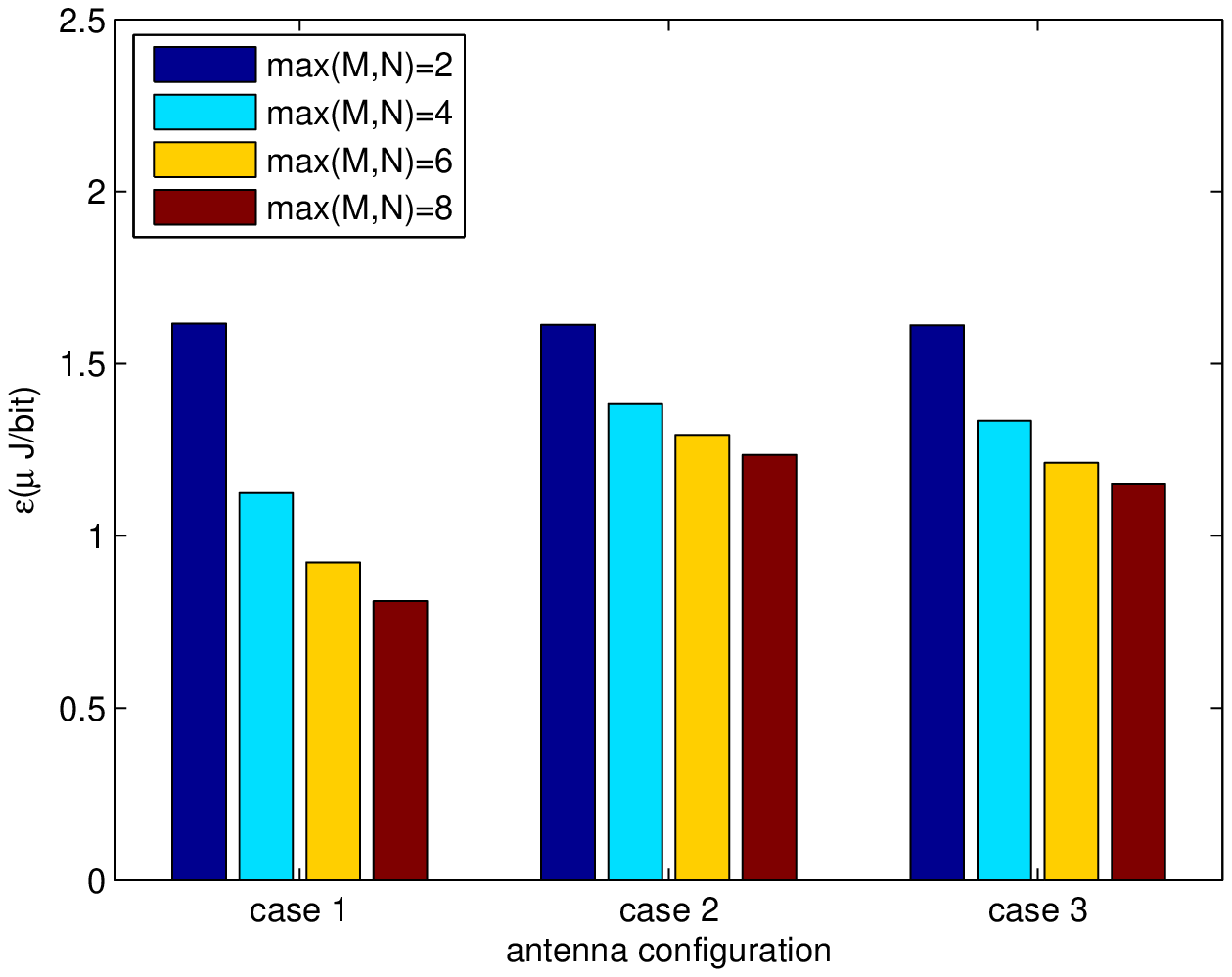}
\caption{Power consumption and EE comparison for different antenna configurations with $d$=50m and $K=32$.}
\label{f4}
\end{figure}

\begin{figure}
\centering
\includegraphics[width=3.5in]{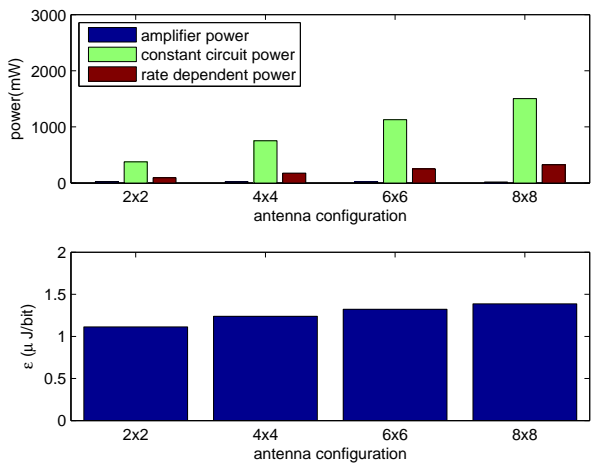}
\caption{Power consumption and EE comparison for different antenna configurations with $d=10$m and $K=32$. $\phi(B\Theta)=\left(B\Theta\right)^{1.2}$ which is nonlinear.}
\label{f5}
\end{figure}
\subsection{Effect of antenna configuration}
Finally, we investigate the impact of antenna configurations on our objective function. Figs. \ref{f3} and \ref{f4} report the value of the objective function at the optimum for $3$ different sets of configurations. In the first set $M$ and $N$  take simultaneously values of $\{2,4,6,8\}$, which is beneficial for the spatial dimensions captured. In the second set $M=2$ and $N \in \{2,4,6,8\}$, which corresponds to receive diversity. The third set corresponds to transmit diversity: $M \in \{2,4,6,8\}$ and $N=2$. As it appears from figure \ref{f3}, for the system parameters selected here and with $d=10$m and $K=32$, set $1$ is more beneficial when the number of antennas increases. This is mainly due to the pre-log factor associated with the spatial multiplexing gain. On the contrary, it is detrimental to only increase $M$ (the number of transmit antennas) or $N$ (the number of receive antennas). This can be explained by the fact that the short distance between transmitter and receiver translates into a small transmit power, which turns out not to be the dominating term in the total power consumption. Otherwise stated, the total power is more impacted by the constant term, meaning that the optimal point corresponds to a high SNR value \cite{BL11}. Hence, when the number of transmit (receive) antennas increases, the diversity gain cannot compensate the increase of power consumption due to the additional RF chains. Fig. \ref{f4} compares the three sets of antenna configurations for a larger distance, i.e. $d=50$m. While the conclusion remains the same for the first set of configurations, increasing the number of antenna turns out to be beneficial for the two other sets, exploiting diversity. As a matter of fact, the transmitted power increases due to distance and becomes dominating in the total power. The optimal operating point corresponds to a lower SNR \cite{BL11} at which diversity gain prevails over the additional power due to the increasing number of RF chains.

Interestingly, our previous conclusions are sensitive to the value of $\alpha$. Fig.\ref{f5}  reports result for a scenario where the $\phi(\cdot)$ function is nonlinear with $\alpha=1.2$. The first set of configurations, i.e. $M=N$, is considered. It now turns out that increasing $M=N$ is detrimental, in opposition to what we had for $\alpha=1$. This is due to the rate dependent power consumption that grows with the multiplexing gain faster than the information rate. This shows the crucial role played by the power consumption model in optimally designing the link. 


\section{Conclusion}
In this paper, we studied the energy efficient design of precoders for point-to-point MIMO OFDM systems. We showed that for the total power made of a constant term plus another one that is increasing and convex with the transmission rate, the consumed energy per bit is a quasi-convex function of the total transmission rate. Thanks to that, the problem of minimizing the consumed energy per bit could be reformulated as a convex fractional program and solved by means of a simple bisection algorithm. The effects of various system parameters on the optimal value of the objective function have been analyzed and illustrated by means of computational results. 

\section*{Acknowledgements}
The authors would like to thank BELSPO for funding the Belgian IAP BESTCOM project.

\end{document}